# The structure of multiplex networks predicts play in economic games and real-world cooperation

Curtis Atkisson[1,*]: Orcid # 0000-0003-3575-6871
Monique Borgerhoff Mulder[1,2]: Orcid #0000-0003-1117-5984

(1) Department of Anthropology, University of California Davis, Davis, CA 95616

(2) Department of Human Behavior, Ecology and Culture, Max Planck Institute, Leipzig, Germany 04103

**\*** corresponding author: PO Box 101, North Hartland, VT 05052; 573-746-1804, atkissoncj@gmail.com

**Abstract**

Explaining why humans cooperate in anonymous contexts is a major goal of human behavioral ecology, cultural evolution, and related fieldsWhat predicts cooperation in anonymous contexts is inconsistent across populations, levels of analysis, and games. For instance, market integration is a key predictor across ethnolinguistic groups but has inconsistent predictive power at the individual level. We adapt an idea from 19th-century sociology: people in societies with greater overlap in ties across domains among community members (Durkheim's "mechanical" solidarity) will cooperate more with their network partners and less in anonymous contexts than people in societies with less overlap ("organic" solidarity). This hypothesis, which can be tested at the individual and community level, assumes that these two types of societies differ in the importance of keeping existing relationships as opposed to recruiting new partners. Using multiplex networks, we test this idea by comparing cooperative tendencies in both anonymous experimental games and real-life communal labor tasks across 9 Makushi villages in Guyana that vary in the degree of within-village overlap. Average overlap in a village predicts both real-world cooperative and anonymous interactions in the predicted direction; individual overlap also has effects in the expected direction. These results reveal a consistent patterning of cooperative tendencies at both individual and local levels and contribute to the debate over the emergence of norms for cooperation among humans. Multiplex overlap can help us understand inconsistencies in previous studies of cooperation in anonymous contexts and is an unexplored dimension with explanatory power at multiple levels of analysis.

**Main Text**

**Introduction**

Across species, cooperation with strangers is rare (Axelrod, 1981). This is because any non-cooperating individual within a group of cooperators receives the benefits of the cooperation of others without incurring the costs (Grossman and Hart, 1980). Hamilton (1964) argued that individuals could increase their inclusive fitness by helping those who are related to them. Subsequently, Trivers (1971) proposed that cooperation could arise through reciprocity, specifically that individuals take turns in helping each other, which can be extended to indirect reciprocity where others return the favor (Nowak and Sigmund, 1998). Recent work has shown that networks, both spatially explicit (Nowak and May, 1992) and not (Ohtsuki et al., 2006), can encourage the persistence of cooperation. Each of these models are predicated on the assumption that individuals choose to cooperate or not because of the specific costs and benefits accrued given the specific ecological and demographic conditions (Apicella and Silk, 2019).

Humans, however, stand out for their prosociality towards strangers (Bowles and Gintis, 2011), which has resulted in two distinct explanatory frameworks. One framework sees cooperation as "driven by environmental differences in demography and ecology" (Lamba and Mace, 2011), which include the general processes above. Specifically, prosocial behavior arises because of individuals making optimal decisions based on a mix of their preferences and current expected patterns of behavior (Gurven, 2004). A separate class of explanations attributes a prominent explanatory role to "norm psychology" (Chudek and Henrich, 2011). In line with the standard definition, norms are behavioral expectations for how individuals should behave in a certain context shared and enforced by a community. Boyd and Richerson (2002) have shown that any behavior can become stabilized as a norm, even if it is costly to the individual. Fundamentally, norms arise because groups that can stabilize on norms of behavior outperform groups that cannot (Nowak 2006; Boyd and Richerson, 2002). This results in a "norm psychology" where individuals are predisposed to identify and follow norms (Chudek and Henrich, 2011). An explanation from norm psychology does not reject any of the general mechanisms given above, it just places more emphasis on the development and enforcement of normative expectations.

Process is often inferred from patterning. Accordingly, to separate these two explanations for why people behave prosocially towards strangers, investigators have examined the patterning of cooperation within and between ethnolinguistic groups, predicated on the following logic: if most of the variability in cooperative behavior is seen at the level of the ethnic group, this indicates the importance of norms in structuring behavior (Henrich et al., 2004), whereas if most of the variability occurs at the individual or local level, this indicates the importance of local socioecological and demographic factors in structuring cooperation (Lamba and Mace, 2011).

The debate over whether group level culturally-transmitted norms or locally-adaptive processes operating at the individual and local level are primarily responsible for human prosociality, that played out in the pages of this and other journals (Lamba and Mace, 2011; Henrich et al., 2012; Lamba and Mace 2012; Henrich, 2011) is problematic for two reasons. First, between ethnic group variation may reflect adaptive processes, and second locally variable behavior can be codified into norms (Tsusaka et al., 2013). Furthermore, despite a large literature showing how variation in game play and/or real world cooperation is patterned at the ethnic (or national) level (e.g., Gachter et al., 2010; Talhelm and English, 2020, Ensminger and Henrich, 2014; Gelfand et al., 2011), our understanding of the predictors of local variation (i.e., variation among communities within ethnic groups) in cooperation, together with predictors of individual variation, is more limited (though see Bigoni et al., 2016; Aswani et al., 2013; Smith et al., 2016; Du and Thomas et al., 2019; Wu et al., 2015; MacNeil and Wozniak, 2018). This lack of a systematic understanding of what drives local level and individual variability in cooperative tendencies is a problem because the patterning of such tendencies is likely to contribute to the emergence and differentiation of norms.

Accordingly, we present an empirical analysis of cooperation within a single ethnolinguistic group, motivated by our suspicion that norms (by which we mean moral expectations structuring behavior, not mere statistical patterning) most likely emerge at multiple levels (not just ethnolinguistic groups). Furthermore, we suspect that individual behavioral adaptations to the costs and benefits of cooperation coevolve with norms (including the costs of punishment) at local levels of social organization, such that pitting explanations of norms and individual behavioral adaptations against each other is unwarranted. Rather, we propose that insofar as norms and locally adaptive behavior coevolve, we need to look for underlying dimensions that can account for both the norm and variation in individual behavior regarding that norm. Accordingly, we introduce a framework dating back to the classical sociologist Emile Durkheim that identifies overlap in an individual's various social networks as an important dimension upon which cooperation will vary (in both anonymous and non-anonymous interactions). Using this framework, we test two predictions employing multilevel modeling and multi-layer networks with new data from the Makushi of Guyana.

*Joint causation of norms and individually variable behavior*

There is an old idea in the social sciences, now implementable using modern network tools, that may shed light on how and why cooperative norms vary, providing an avenue for tackling the question of why humans are so variable in their willingness to cooperate with strangers. Durkheim (1893) proposed that societies are structured in two distinct ways. In societies with "mechanical" solidarity people do most of their activities (e.g., work, play, pray) with the same set of people, and therefore have high overlap. In contrast, people in societies with "organic" solidarity work, play and pray with distinct sets of people, and therefore have low overlap. We adopt, modernize, and adapt this approach here for, to the best of our knowledge, the first time at both the individual and community levels.

With this distinction in mind, we propose that people in societies with high overlap have important network partners with whom they interact across many domains, the loss of which would cost them dearly. Accordingly, they are likely to pay to maintain those important relationships (which might include engaging in costly cooperation with people with whom they currently have relationships). They are also likely to place less value on recruiting new network partners. As such, in societies with "mechanical" solidarity we expect behavior to be driven by local norms favoring cooperation with community members and caution with anonymous others, who may be strangers.

Conversely, individuals in societies with low overlap place less importance on any specific relationship because losing a relationship with another individual does not jeopardize ties across many domains. In these societies with "organic" solidarity, any single relationship can be broken with less consequence, such that individuals will be less willing to pay high costs to maintain relationships. In such contexts social networks are likely to be in greater flux at any point in time and recruiting new network partners becomes more important. One way people who need new partners could signal their value as a possible cooperative partner is showing their willingness to share resources with other people with whom they do not currently have a relationship, such as in an anonymous context. This could potentially lead to a norm of high levels of sharing with people who are not currently in one's network, a strategy that might be stable if there are sufficient benefits from recruiting short-term network partners.

As noted above, previous work that examines the patterning of cooperation using economic games has been more successful at explaining variation between ethnic groups than the variability seen between communities or at the individual level. It has also revealed considerable inconsistency between findings (see Henrich, et al., 2004 [especially compared to Ensminger and Henrich, 2014] and Gurven, et al., 2008). The Durkheimian framework of network overlap we propose here generates the following hypotheses to be tested with both experimental game play and real-world cooperation.

> **H1: Individuals with high overlap within their villages and individuals in villages with high overlap will contribute less in anonymous interactions**. We expect these intra and inter-community patterns because individuals and villages that have high overlap have network partners that are stable because they are so enmeshed; as such, we expect them to have a lower need to signal the quality of their partnership to potential new partners.
>
> **H2: Individuals with high overlap within their villages and individuals in villages with high overlap will engage in more real-world cooperative events**. Since individuals and communities with high overlap are so enmeshed, losing a partner would be a significant event; as such, we expect them to have a greater impetus to maintain their current relationships.

Durkheim's (1893) framework emphasizes the overlap of partners across domains, requiring us to keep domains separate from each other. Keeping domains separate from each other results in a multi-layer network, of which the analysis has only recently begun (Kivelä et al., 2014; Finn et al., 2019; Atkisson et al., 2020). By keeping domains separate, we can look at the overlap of an individual's partners across those domains. This allows us to calculate a value for each nested level of interest, which does not require us to simply dichotomize societies as Durkheim did.

*Testing these hypotheses*

The North Rupununi area of Guyana provides a good opportunity to test these ideas. Most people living in this area are Makushi and practice swidden horticulture with foraging. Most of their animal protein comes from fish, with occasional supplements of hunted meat. For their livelihood the Makushi hunt, fish, forage, and garden on their own when they can, calling large groups together to help them accomplish things they could not do on their own (*mayu*, see Elias, Rival, and McKey, 2000; Schacht, 2015), such as clearing land for cultivation or thatching a new house (see Materials and Methods). A road was recently built that opened the area to wage labor as well as short- and long-term labor migration. When people change their subsistence strategies, social ties in traditional subsistence strategy domains typically erode, requiring their erstwhile partners to find new partners in the affected domains. This has led to individual variation in network reformation and, as different communities have become differently dependent on these forms of labor, average network overlap at the village level varies. See Materials and Methods for a more complete description.

We (CA) gathered data using a stratified random sample of people chosen randomly from 9 villages chosen to maximize variation on three main variables (see SM Table 1 in S2 in SI; one village will not be included in this analysis, see Materials and Methods and S4 in SI; village specific sample size range 26-30 participants). To measure our key predictive variable, overlap, we asked our respondents to whom they lent (or gave), and from whom they borrowed (or received) 52 distinct items (e.g., hunted meat, generator, spade, etc.; S3 in SI for all survey questionnaire items used in this analysis). Each item in each direction represents a domain of interaction. We also include who they did four types of hunting with. *Individual overlap* is calculated as the proportion of ties that are present in more than one domain. *Village overlap* is the mean of the overlap of everyone in the village. While village overlap is necessarily related to village size, the size of these villages (minimum 151) means this dependency is negligible.

We measured prosociality in three different ways. First, we played the Dictator and Ultimatum Games to assess how cooperative people are in anonymous contexts. In the Dictator Game (DG), participants were asked to contribute to an unknown member of their community from an allowance of 1,000 GYD (representing ⅓ of a day's wages). Second, we used the Ultimatum Game (UG). In this task participants were asked to make an offer to a different unknown member of their community, which their partner could reject, from an allowance of 1,000 GYD. The naïve economically rational choice in the UG is to give the smallest non-zero amount of money whereas the naïve economically rational choice in the DG is zero, allowing us to examine how the effect of overlap in anonymous situations changes given the difference in rational choice. Finally, we also collected data on how often respondents participate in a real-world cooperative activity (*mayu)* to assess their level of intensive cooperation with network partners.

In general, we are interested in predicting

$$\pi_{i,V} = a_V + B_1 overlap_i + B_2 overlap_V \tag{1}$$

where $\pi_{i,V}$ is the response of individual *i* in group *V*, and $a_V$ is a random effect for each village. The economic games have a distribution requiring a sequentially ordered categorical model while the count of times someone engages in *mayu* is modeled using a negative binomial distribution (see Materials and Methods).

**Results**

In both DG and UG there is a modal offer at 200 GYD (20%), though 300 GYD is also a mode in the UG (see Figure 1). Within the span of 100 to 400 GYD, DG offers look like they could be normally distributed but there are large outliers at 0 and 500 GYD. An offer of 0 GYD is the economically rational choice in the DG, while the economically rational choice in the UG is to give the smallest non-zero amount possible (assuming one's partner is economically rational). This is why we see large spikes at 0 GYD for the DG and almost no offers of 0 in the UG. Anything above those offers (0 GYD for the DG and 100 GYD for the UG) means that people are using some other decision criteria than strict economic rationality. To the extent that humans have an aversion to inequality (Fehr and Schmidt, 1999), we would expect them to give 500 GYD—and we see a large increase in the number of offers of 500 GYD, at least in the DG.

To test the first hypothesis, offers in the anonymous DG and UG were examined in relation to individual overlap and village mean overlap, in a model containing a random effect for village (Table 1). The probability of offering 0 GYD in the anonymous DG increases as overlap increases while the probability of giving more than 200 GYD drops (Fig 2). The effects of both village-level and individual overlap on offers to anonymous recipients in the DG are negative, while the effect of overlap on offers to anonymous recipients in the UG was in the same direction, though not clearly different from zero. The variation organized at the village level in a model of DG with no covariates is 3.8%% while the variation organized at the village level in a model of DG with the covariates included is 2.2%. The same numbers for the UG are 7.9% and 6.8%, respectively. These estimates of variation organized at the village level are likely underestimates (see S6 in SI for complete discussion) but fall within the range of estimates of variation organized at the community level from Lamba and Mace (2011) and Henrich and colleagues (2012).

To test the second hypothesis, the number of times an individual engaged in *mayu* was predicted by village- and individual-level overlap, again with a model including village random terms. The effects of both individual and community overlap are positive, so that people with higher overlap engage more in *mayu* and people who live in communities with a high average overlap engage more in *mayu* (see Table 1; see S5 in SI for an analysis that shows lack of sensitivity to outliers). Across the range of individual overlap, we predict participation in an additional 23 *mayu* events each year (Fig 3). Across the range of community overlap, we predict an additional 13 *mayu* events engaged in each year (Fig 4). The variation organized at the village level in a model with no covariates is 41% while the variation organized at the village level in a model with the covariates included is 1.8% (standard deviation of village-level means without covariates = 2.97, standard deviation of village-level means with covariates = .49). When community size is included as a variable in any model (i.e., adding $B_3 size_V$ to equation 1), it has no effect on any outcome variable.

**Discussion**

These results make two contributions to the literature on the evolution of prosociality: first, they validate the significance of Durkheim's concepts of mechanical and organic solidarity for structuring cooperative tendencies, insofar as the extent of network overlap affects the patterning of cooperation with both known community members and strangers; second, they demonstrate the importance of individual and community level factors in shaping cooperation. People in villages with a higher average overlap have lower contributions in anonymous interactions as captured through experimental game play and report more participation in real world cooperative labor groups than people in villages with a lower average overlap. Within villages, people with more overlap have lower contributions in anonymous game play settings and higher reported participation in cooperative labor groups, also consistent with the hypotheses. While it is difficult to elicit the existence of norms *per se*, or to demonstrate them through experimentation, the patterning of these finding is consistent with expectations that norms and individual strategizing impact both anonymous and non-anonymous cooperation as a function of network overlap.

Our findings have implications for the debate over the saliency of cultural norms versus individual and local adaptation in contributing to the emergence of cooperative norms (Lamba and Mace, 2011; Henrich et al., 2012). We posit that each of these positions is incomplete, and that norms will emerge at multiple levels as a result of an individual's interactions, leading to a certain set of expectations. Insofar as most people's interactions in small-scale societies will be within a small relatively closed group, their expectations will be the same, which, upon observation, we would call a norm.

Norms result from the outcome of a subtle dyadic interplay between individual self-interest, the benefit of coordinating behavior with others, and the costs of norm breaking (e.g., Glowacki, 2020). Some individuals in some circumstances may benefit more (or suffer less) from breaking the norm than following it. In our work here we have not demonstrated these dynamics, but we have shown how different patterns of cooperation emerge at local levels, as we hypothesized in response to the extent of network overlap. This apparent facultative variability in tendencies towards cooperation, both in the real world and in anonymous interactions, has the potential to sow the seeds for change in local norms. From thence, and becoming ever more speculative, we suggest that because norms are inherently sticky in species where the social ecology comprises much of the adaptive landscape (Richerson and Boyd, 2008), the fission-fusion dynamics of human populations will lead to daughter populations inheriting many of their norms from their parent population. In other words, norms may become ossified at higher levels of organization, thereby characterizing an ethnolinguistic group. In this sense, both the arguments of Lamba and Mace and of Henrich and colleagues have value.

Our results show that one dimension considered at both the individual and community level can influence cooperative behavior. This shows how locally adaptive norms can take root in a community. While Henrich and colleagues' extended response (2011) to Lamba and Mace (2011) provides some guidance, in our view it minimizes the key role of individual circumstance/behavior and strong dependence on initial or small changes in conditions. This also highlights the fact that norms such as prosociality expected in anonymous interactions arise at the locally salient level. In our study this is the village, but it is reasonable to expect that different levels will become salient in differing contexts, as shown in the comparison of rainfed and irrigated farming areas in the Philippines. Tsusaka et al (2015) found that when the salient cooperative group became more geographically constrained (by the need to manage irrigation), people tended to behave more similarly to their neighbors in economic games. Not only that, but punishment subsequently resulted in much more cooperation among farmers in the irrigated compared to rainfed zones. In this case, the locally salient context where the norm is arising is dependent on the local agricultural technology.

Our focus on how network overlap patterns cooperation at both individual and community levels resolves some of the inconsistencies previously detected in the literature. We suggest that some of the independent variables reported to predict behavior in anonymous economic games might also be associated with differences in network overlap, and that these differences affect cooperation in the expected direction. These include larger contributions from Hadza living in larger camps and with fewer siblings giving more (Marlowe, 2004), Zimbabweans who live in communities that had been forcibly resettled (Barr, 2004), New Guineans with more cocoa fields and increased wage labor (Tracer, 2004), higher income Orma (Ensminger, 2004), and Pahari Korwa who are older and have more connections outside of their communities (Lamba and Mace, 2011). Similarly, smaller contributions come from married Tsimane (Gurven et al., 2008) and Pahari Korwa with more sisters in their community (Lamba and Mace, 2011). For further possible consistencies with our framework, see S7 in SI.

A key strength of Henrich and colleagues' (2004, 2014) approach lies in their ability to identify covariates that predict variation across ethnolinguistic groups. The present study complements previous work by showing how network overlap can serve as a unifying dimension that helps us understand variation in cooperative behavior at both community and individual levels. These ideas should be further tested using multiplex methods across different societies. By merging a classical formulation of how societies differ in features of their multiplex networks with predictions at the individual level this paper brings a new framework to studying what shapes variation in prosociality among humans. This helps us understand why patterns of cooperation might change as communities become more linked to the outside world.

**Materials and Methods**

*Makushi*

The Makushi are the largest extant group from the Carib language family, with over 42,000 people who claim membership (Instituto Socioambiental, 2018), 9,000 of which live in Region 9 of Guyana. Most communities have access to a town of ~1,700 people via a 3-5 hour mini-bus ride, though the route is often impassable in the rainy season, though some communities require over a day's travel. Makushi subsistence activities consist of both cooperative and individualistic endeavors and include swidden horticulture and foraging. The other main subsistence activity is fishing, supplemented with hunting occasionally. Nearly all hunting is done with bow and arrow, with fewer than 10 licensed firearms in the entire region and very few unlicensed firearms. The recent construction of a road through the area has opened up short- and long-term labor migration as well as wage labor, which, because of access to the road, led to people in some communities needing to restructure their subsistence networks more than in other communities. One primary traditional institution that remains is that of *mayu*, which is a cooperative labor event. See S1 in SI for further details on the ethnographic context.

*Sampling frame*

Because of the specific predictions for within- and between-community cooperation derived from our hypotheses, our research design required the sampling of both communities and individuals within these communities (Schacht and Borgerhoff Mulder, 2015). To maximize variation in community-level variables, we purposefully selected communities depending on their population size, number of government workers (a proxy for wage labor), and distance from the administrative center of the region (towns to the west are closer to employment near the Brazillian border). Communities were categorized into high or low values on each of those variables and were selected to cover nearly all combinations of high and low values on all variables (see S2 in SI). One village was dropped from the analyses reported here because of a significant disruption in the network, see S4 in SI for an explanation and analysis with that village included.

Within villages, researchers worked with Community Health Workers and the Village Council to obtain a census of all people who currently call the village home, whether or not they were present. From this census, 30 households were randomly selected and one adult from each selected household was randomly selected. If a selected participant was not going to be in the village during the 30-day study period, they were replaced choosing randomly from those houses which had not yet been selected (but also including the household to be replaced) in the manner above. While this will result in households with migrant laborers being represented at lower rates, replacing them with a person in the household who remained would have biased the entire sample.

*Interview*

The interview began with demographic and socioeconomic questions, then respondents played economic games, then respondents answered the network questions, and finally the respondents were asked about their participation in *mayu*. If participants were unable to report monthly participation, their estimates of yearly rates were used; all responses were transformed into yearly rates.

*Economic games*

Two economic games were played with each participant, the Dictator's Game (DG) and Ultimatum Game (UG; Henrich et al., 2004). In both games, the focal individual was given 1,000 GYD (⅓ of a day's wage) and was told that they were being partnered with a random person from the village. They were told that they would not know who their partner was and that their partner would never find out who they were.

After the script for the DG was read, the research assistant gave 4 examples. After the script for the UG was read, the research assistant gave examples with 4 different amounts of money, with the hypothetical second player both rejecting and accepting the offer. The respondent was then tested on their knowledge. If the respondent answered incorrectly, the research assistant would tell them they were wrong and explain the correct answer. Respondents continued in testing until they answered three test questions in a row correctly (modal number of repeats required was 0). One respondent was removed from the sample for inability to understand these games.

*Network and network measures*

Participants were asked about other people they do things with across many domains of interaction using a culturally appropriate name generator (S3 in SI). For this study, all 54 of the network domains regarding from whom someone gets/borrows or to whom someone gives/lends something were included. This could include anything from foodstuffs to phone credit to large equipment such as chainsaws.

As this was not a complete network, this forms a directed, multiplex ego-network (Wasserman and Faust, 1994; Kivelä, et al., 2014). Each individual had the same potential alters they could nominate (i.e., everyone in the area) but only nominated a subset of them in each domain. Each domain forms a layer of the multiplex network.

Overlap was calculated by dividing the number of interactions across all partners in all domains that someone had with a partner who was represented in more than one domain by the total number of interactions across all partners in all domains. If an individual had 56 partners across all domains and 19 of those were with someone mentioned more than once, their overlap would be 19/56 = 0.339. Village overlap was calculated as the mean of the overlap for all individuals in the village.

*Statistical analyses*

We are interested in predicting the three outcome variables, number of times per year someone engages in *mayu* and contribution in the two economic games, using the same variables. Specifically, we are interested in the effect of overlap and village overlap. Since we are explicitly assuming and testing if norms arise at the village level, this is a multi-level (or random effects) model. For simplicity, we can present the equation of interest for all outcome variables as in equation 1.

Since respondents in the economic games could only select between 0 and 1000 GYD in 100 GYD increments, the offer is best modeled as an ordered logistic regression (Jackman, 2009). The number of times per year someone participated in *mayu* is a count variable with unequal mean and variance, necessitating the use of a negative binomial distribution (Cameron and Trivedi, 1998).

We estimate the parameters of interest in these models using Hamiltonian Monte Carlo with No U-Turn Sampling as implemented in Stan (Carpenter, et al., 2017) in R (R Core Team, 2017) using the convenience provided by the package brms (Bürkner, 2017). Since this is a fully Bayesian method of estimation, we do not report p-values but instead 89% credible intervals (McElreath, 2016).

We calculate the proportion of variance organized at the village level in a model only including random effects and the proportion of the variance organized at the village level in a model including covariates using the intra-class correlation (ICC) for each of our models of interest. Further technical details regarding the statistical analyses may be found in S6 in SI.

**Acknowledgements and funding sources**
We would like to thank Cristina Moya and Andy Sih for extensive comments on previous drafts of this paper. Richard McElreath, Pete Richerson, and Fushing Hsieh provided feedback on the design of the project. Mark Grote and Damien Caillaud provided statistical advice. The EEHBC lab group at UC Davis was invaluable. Two anonymous reviewers of an NSF DDRIG application provided valuable guidance. Any remaining errors are our own. We are thankful to Hon. Sydney Allicock who supported this research with his approval of the project for the Ministry of Amerindian Affairs (now the Ministry of Indigenous Peoples' Affairs) in Guyana. Rebecca Xavier and Ricky Moses gathered data in the field. We are especially grateful for the warmth and easy participation of the people of the North Rupununi area of Guyana. Funding for this work was provided by the National Science Foundation (#1558890; Monique Borgerhoff Mulder PI, Curtis Atkisson co-PI) and The Wenner-Gren Foundation (Dissertation Fieldwork Grant; Curtis Atkisson PI).

**Data availability**
All data and code will be made publicly accessible via github upon acceptance of the paper. Code and data have been provided separately for review purposes.

**Author contributions**
CA conceived of and designed the study, gathered data, performed statistical analyses, and wrote the paper. MBM assisted with both the design of the study and the writing of the paper.

# Tables

| | Table 1: Effect of overlap on cooperation across 8 villages | | | | | | | | |
|---|---|---|---|---|---|---|---|---|---|
| | *Dictator Game offers* | | | *Ultimatum Game offers* | | | *Mayu participation* | | |
| **Effect** | **Estimate** | **L 89% CI** | **U 89% CI** | **Estimate** | **L 89% CI** | **U 89% CI** | **Estimate** | **L 89% CI** | **U 89% CI** |
| Individual overlap | -2.83 | -5.41 | -0.25 | -1.16 | -3.68 | 1.38 | 2.53 | 0.11 | 5.09 |
| Village overlap | -23.10 | -43.61 | -2.50 | -18.77 | -44.88 | 7.14 | 24.35 | 20.82 | 27.61 |
| Village random effect: std. deviation | 0.27 | 0.02 | 0.66 | 0.49 | 0.10 | 1.00 | 0.49 | 0.10 | 0.95 |

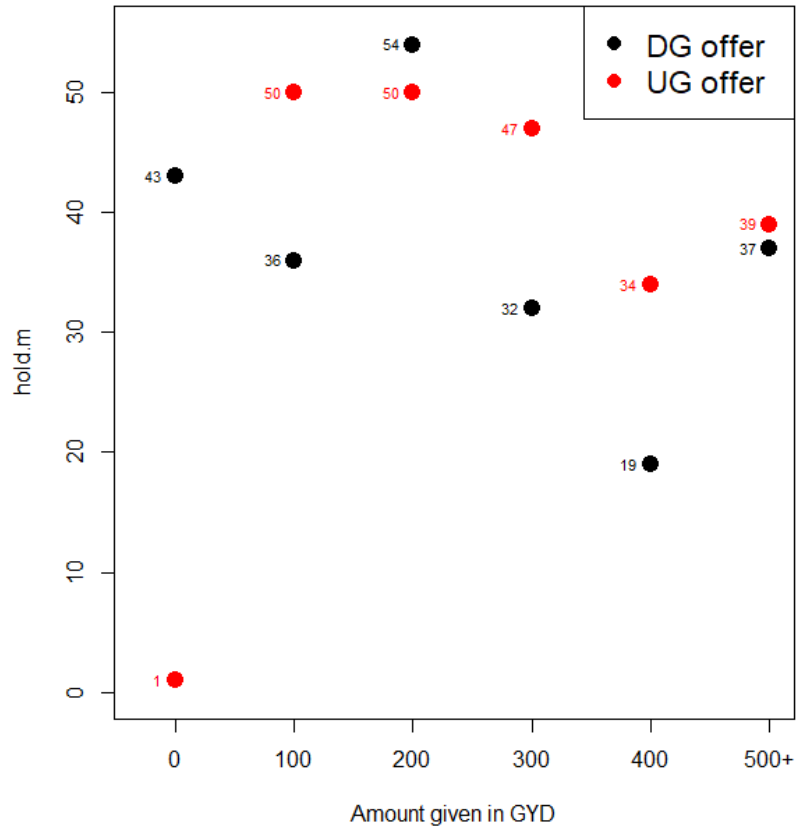

**Figure 1: Number of people giving and offering each amount.** shows the number of people in the sample giving or offering each of the values (0 to 500 GYD) for both the Dictator Game (DG; black) and Ultimatum Game (UG; red)

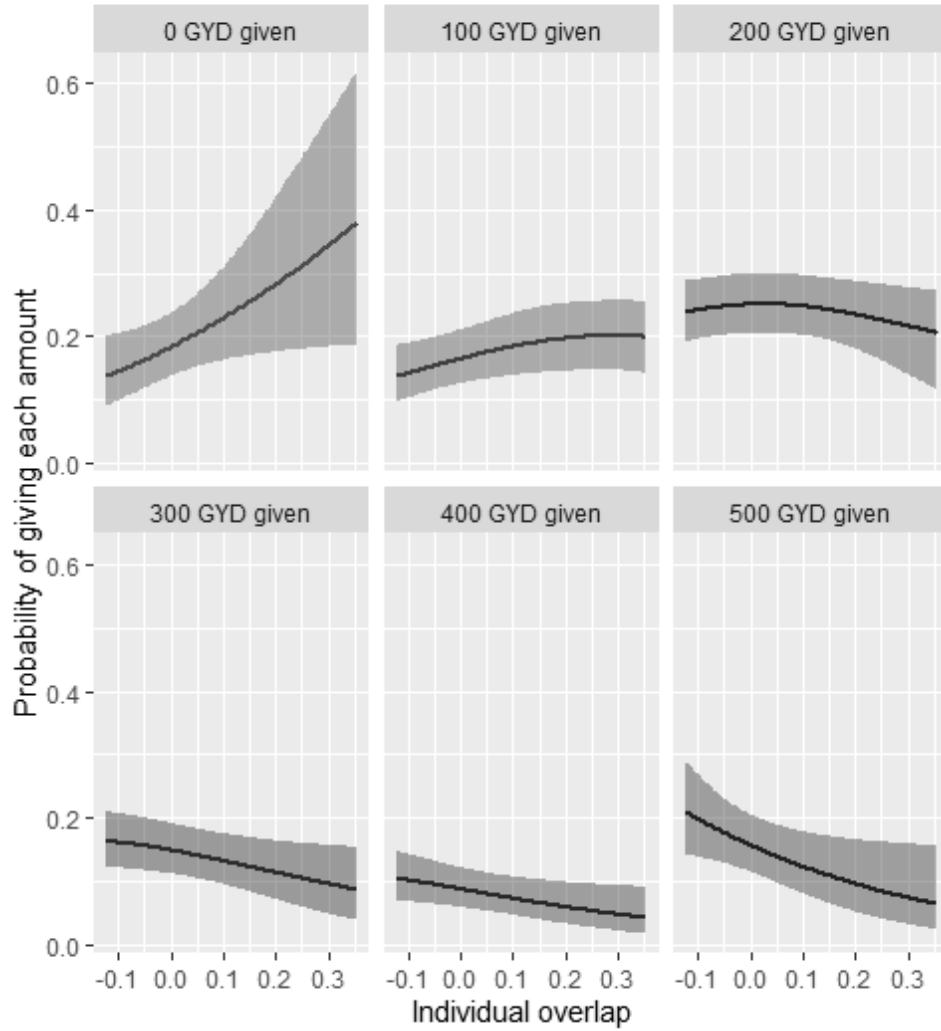

**Figure 2: Probability of giving each amount by individual overlap.** This figure shows the probability of an individual giving each amount of money as their overlap increases. Amount given ranges from 0 to 500 GYD in 100 GYD increments and the probability of giving each amount across the entire range of individual overlap is separated into its own panel. The estimated probability is the solid line and the shaded areas represent the 89% credible intervals around the estimate. As overlap increases, probability of giving 0 GYD increases dramatically and probability of giving more than 200 GYD approaches 0.

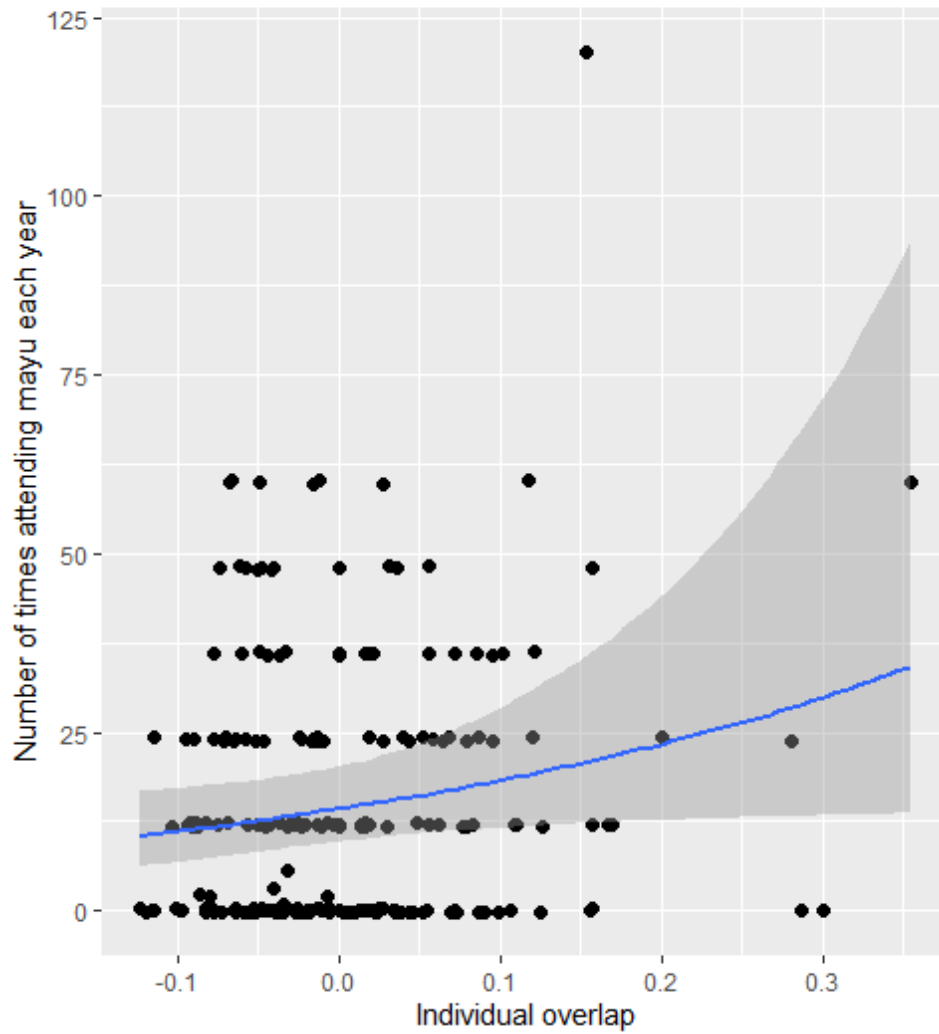

**Figure 3: Marginal effect of individual overlap on number of times attending mayu in a year.** This shows the estimated marginal effect of individual overlap on the number of times in a year an individual attends mayu. The line shows the mean and the shaded areas show the 89% credible intervals. The points are the original data (N=226).

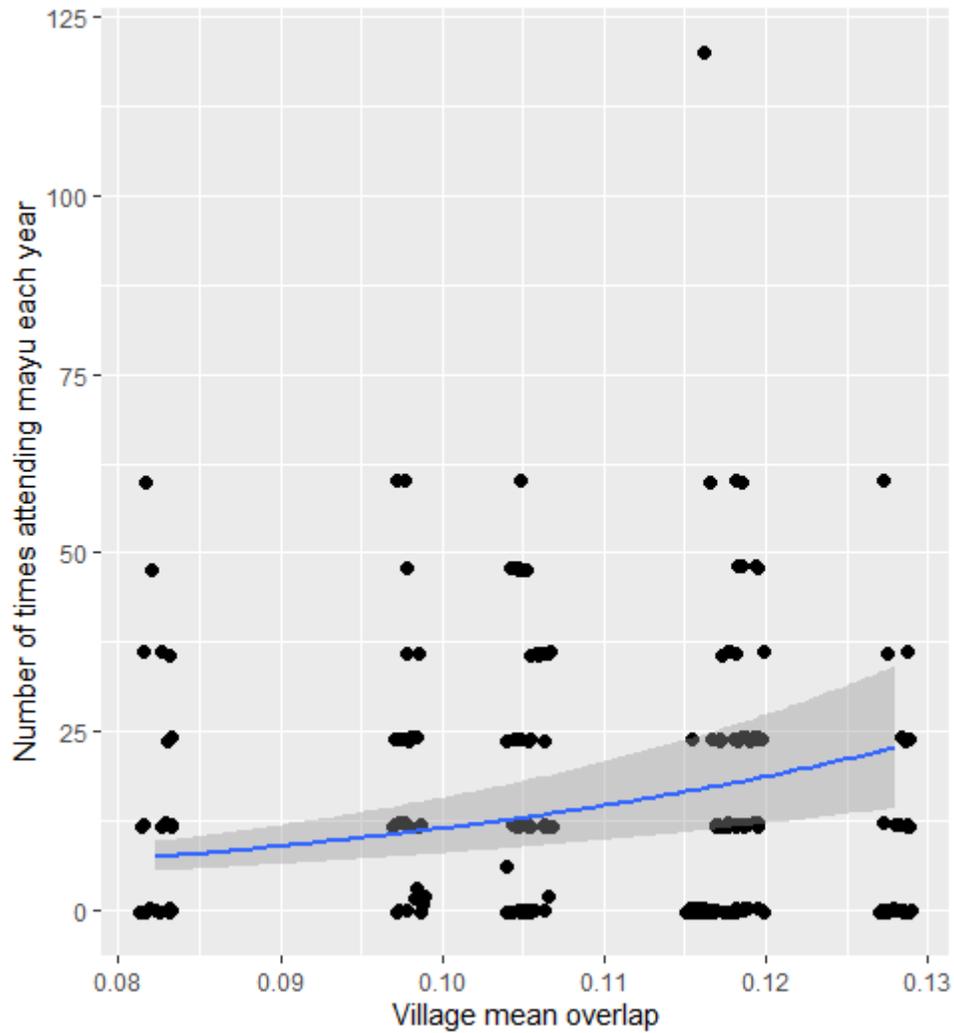

**Figure 4: Marginal effect of mean village overlap on number of times attending mayu in a year.** This figure shows the estimated marginal effect of mean village overlap on the number of times in a year an individual attends mayu. The line shows the mean and the shaded areas show the 89% credible intervals. The points are the original data (N=226).

**Supplementary Information**

This is the supplementary information appendix to "The structure of multiplex networks predicts play in economic games and real-world cooperation" by Atkisson and Borgerhoff Mulder.

**S1: Ethnographic background to the study site**

To meet the design criterion of this study, we chose to work in a single ethnolinguistic group living across a spatially and socially heterogeneous environment. As an ethnolinguistic group, the Makushi are spread between Brazil and Guyana, with limited groups living in Venezuela. Many Brazilian Makushi only speak Portuguese, while most Guyanese Makushi speak both English and Makush-pe (the Makushi name for their language). Approximately 9,000 Makushi live in Region 9 of Guyana (Upper Takatu-Upper Essequibo) between the Rupununi River (a tributary to the Essequibo River; a part of the Rupununi portal that connects the Amazon and Essequibo drainages [de Souza, et al., 2012]) and the Pakaraimas mountains. Most communities in North Rupununi have access to Lethem at the Brazilian border via a 3-5 hour mini-bus ride, though the route is often impassable in the rainy season. Some communities in this area can require over a day's travel to arrive in Lethem.

Makushi subsistence activities consist of both cooperative and individualistic endeavors and include swidden horticulture and foraging. They cut fields of one to three acres at the foot of the mountains and plant mainly cassava (*Manihot esculenta*), watermelon (*Citrullus lanatus*), eddoe (*Colocasia antiquorum*), and various species of pepper. They also plant other crops as desired, such as tobacco. The second main food is fish, which they get from the Rupununi river. Traditionally, Makushi would spear or dive for fish hiding between rocks. This has given way to poles and, especially, gill nets that increasingly catch smaller and smaller fish. The Makushi describe gill nets according to how many fingers can be inserted into a hole. In the past, it would be rare to see nets smaller than three- or four-finger, now one- and two-finger nets are common. The Makushi fish for dozens of species. Makushi also hunt. Hunting primarily takes place during the rainy season when animals have been driven to hilltops to avoid the floods. Hunting may also take place when someone spots a herd of peccary moving across the savannah. Many people will be recruited for this "hog shout" and tens of animals can be killed. This is rare. Makushi opportunistically hunt large rodents such as agouti (genus *Dasyprocta*) around their farms either upon encounter in the day, or intentionally during the night if animals are ruining their crops. Finally, some hunting of larger game such as deer and tapir takes place during the night, though this is dangerous because of the deadly pit vipers (genus *Lachesis*) in the area. Nearly all hunting is done with bow and arrow, with fewer than 10 licensed firearms in the entire region (and very few unlicensed firearms; illegal firearm possession results in a long-term sentence in the prison in Georgetown, known as "life threatening"). Of the few intentional daytime hunts CA has observed, all took place by individuals who have or a team with an individual who has a shotgun.

In order to test the effect of differences in network overlap we took advantage of the fact that Makushi villages across the Rupununi are differentially impacted by migration and changes to subsistence strategies. There have been three primary colonialist or national interventions into the area, the most recent of which was the construction of a road that connects Lethem to the capital city of Georgetown and runs directly through Makushi territory in the North Rupununi (MacDonald, 2014; Torrado, 2007). The road made it possible for some people to switch from traditional subsistence strategies to those based on income. These income opportunities include

short- or long-term shop work in Lethem, (the border town with Brazil), long-term work in Brazil (including on watermelon farms, in ice factories, etc.), and short-term work at the mines and logging camps spread through Guyana. Some villages lie directly on the road, providing easier access to these opportunities (See Fig 1 in Supplementary Materials for a map). Some communities are closer to Lethem while some are closer to the mining and logging camps to the northeast. These differences impact how many people in each community have switched from traditional to cash-based livelihoods, and how long they are removed from traditional subsistence networks (pers obs.). Individuals do not usually provide remittances while they are away working.

*Mayu* is an easily measurable and culturally salient indicator of people's tendencies to cooperate with other in the community. *Mayu* is a combination of hard work and a party. People will typically call *mayu* to accomplish a large task such as cutting a new farm or putting new leaves on a house. Preparations for this event typically begin a few weeks beforehand with the collection of additional cassava. This cassava is mashed and formed into cakes, which are then sprinkled with crushed leaves from the cassava plant, covered with banana leaves, and stored in a dark place for at least several days. The leaves contain yeast on them, which results in a hearty fermented beverage (*kari*). The person calling *mayu* will make enough of this for around 10 people. A day or two before *mayu*, people within that person's close network will be asked to participate. A turnout of any fewer than 6 people would be considered inadequate, with a goal of around 10 people. A little of the *kari* will be offered to workers prior (generally seen as incurring a debt) and during the event, though people are expected to work hard for 6-8 hours and get drunk after everything is done (after being stung multiple times by a scorpion, CA was told that is what happens when one is lazy and stands around the *kari* bowl).

## S2: Community characteristics of selected communities

| Table S+1: Community characteristics of selected communities | | | |
|---|---|---|---|
| *Name* | *Population* | *Government workers* | *Miles from Lethem* |
| 1 | 271 | 5 | 156 |
| 2 | 226 | 6 | 75 |
| 3* | 536 | 7 | 87 |
| 4 | 492 | 43 | 96 |
| 5* | 271 | 6 | 165 |
| 6 | 692 | 6 | 71 |
| 7 | 160 | 2 | 73 |
| 8 | 372 | 10 | 104 |
| 9 | 792 | 7 | 56 |

\* - Riparian village. All others are savannah villages. Shaded are villages >400 people, >10 government workers, and to the east of the administrative center.

**S3: Survey questionnaire items used in this analysis**

Data elicitation for the three individual-level variables used in this study was nested inside a larger questionnaire. This section presents the method of data elicitation for each variable in the order in which it was asked in the questionnaire.

*Economic games*

The script for the DG is as follows (please note that these scripts are written in an English language dialect particular to this area):
*"You will be playing a money game. It's quite simple, it's easy. The game will be played with a person right within the community. You might know the person, but you will not know that you are playing with this person. And the person will not know who they are playing with. You will be given a thousand dollar. You can give to your partner any amount you wish to, from the thousand dollar. The money you give them belongs to them. The money you keep belongs to you."*

The script for the UG is as follows:
*"Just like the first game, we would be playing an offer game. The game will be played with a person right within the community. You might know the person, but you will not know that you are playing with this person. And the person will not know who they are playing with. It will be a different person from the person you played with in the first game, a next person. You would be given a thousand dollars. From that thousand, you can offer your partner any amount you want to offer. If your partner say yes to your offer, they own the money you offer them. And then you own the rest that you keep for yourself. But if they say no to your offer, they gain nothing and you gain nothing. So, if they say no to your offer, you getting 0 dollars, and they getting 0 dollars."*

After this, they were quizzed on how much money each player would get and they needed to answer three questions in a row correctly to move on to playing the game. If they answered a question incorrectly, the correct answer was explained to them and the researcher continued quizzing until three questions in a row were answered correctly.

*Culturally appropriate name generator and domains*

For each domain, individuals were asked (for example), "From whom do you get X?" The person could nominate as many alters as they would like. After each nominated alter, the interviewer would say, "Next one," which continued until the respondent indicated there were no more. At that point, the respondents were asked, "No one else?" This process was repeated until the respondent indicated that there was no one else. No questions were asked about the alters, though any relationship mentioned by the respondent was noted. "Get" and "Give" would be understood in this context to include outright giving (e.g., a cup of sugar) and borrowing (e.g., someone lending a chainsaw for a day).

- Cooking gas
- Vehicle fuel
- Kari
- Salt
- Soap
- Match/lighter
- Bow and arrow
- Transportation
- Carpentry tools
- Farming tools
- Warashi
- Pack animals
- Electrical generator
- Boat or boat motor
- Suitcase
- Phone
- Electricity/inverter
- Phone credit
- Cassava bread
- Parched cassava
- Fish
- Wild meat
- Cooking oil
- Rice
- Purchased meat
- Sugar
- Hunt during the day
- Hunt during the night
- Hog shouts
- Multi-day hunts

*Mayu*

Have you ever attended another person's mayu?
In one month, how many times will you go to mayu someone else calls?
[If 0 to the second question] In one year, how many times will you go to mayu someone else calls?

## S4: Removed village justification and analysis with it included

One village was dropped from all current analyses. This village has a small population, has few government workers, and is far from the border with Brazil. Indeed, the trip from this village can take several days in certain conditions. Since the initiation of the research, a community member who is married to an American established a sports fishing lodge in the area. This lodge is on a private concession (meaning the village does not benefit and only a select group of villagers do). The household running that lodge is from a part of the village that is separated from the rest of the village and in which almost everyone is closely related. As such, the people in that area dramatically increased their exposure to foreigners and decreased their exposure to other members of the village. It seems unlikely that the network measures we use in this study will capture the true network structure for these people. The estimated effects of all predictors decrease and we have less confidence in the effect when we include this village but only the effect of village overlap on DG offers seems to disappear completely. The study was originally designed with this combination of village-level variables oversampled, so there is a community that is included in all the analyses that has a small population, few government workers, and is far from Lethem.

| | Table S2: Effect of overlap on cooperation across 9 villages | | | | | | | | |
|---|---|---|---|---|---|---|---|---|---|
| | *Dictator Game* | | | *Ultimatum Game* | | | *Mayu participation* | | |
| Effect | Estimate | L 89% CI | U 89% CI | Estimate | L 89% CI | U 89% CI | Estimate | L 89% CI | U 89% CI |
| Individual overlap | -2.09 | -4.48 | 0.28 | -1.47 | -3.75 | 0.81 | 2.36 | 0.10 | 4.79 |
| Village overlap | -0.25 | -22.60 | 22.00 | -3.75 | -26.53 | 0.81 | 23.72 | 20.75 | 26.56 |
| Village random effect: std. deviation | 0.57 | 0.21 | 1.05 | 0.59 | 0.24 | 1.09 | 0.51 | 0.18 | 0.92 |

## S5: Analysis of outliers

An analysis of potential outliers was conducted for the analysis by examining the Pareto-k values after Leave One Out cross validation (loo; Vehtari et al., 2017). This indicated that there was a single outlier that was having a strong impact on the estimated effects for the model of participation in *mayu*. Upon removing that individual outlier, the effect of individual overlap substantially overlapped 0, indicating that there is no effect of individual overlap on participation in *mayu*.

It is not appropriate, however, to merely remove this datapoint because it arises in the context of the community this person is in. This person's community forbids the consumption of alcohol. The foundation of *mayu* is an exchange of a token good for labor. Most often, this token is a local alcoholic beverage made from fermented cassava. Because of this, people in this village do not participate in *mayu*. Indeed, the sample from this community had 26 people who never participate in *mayu* compared to a mean of 8.75 people for the rest of the villages (range 5-14 in the other villages). Whereas this person engaged in *mayu* the most in the entire sample and does so by going to nearby villages.

When this entire village is removed from the analysis, the effect of individual overlap is 1.83 (89% CI -0.71–4.63), once again showing that individual overlap increases participation in *mayu*.

## S6: Technical details of statistical analysis

Since respondents in the economic games could only select between 0 and 1000 GYD in 100 GYD increments, the offer is best modeled as an ordered logistic regression (Jackman, 2009). This model uses a logit link function to connect a linear predictor to the probability of being in or below a category in the outcome variable. It estimates parameters for the following
$$Pr(y_{i,V} \leq c) = F(\tau_c - \pi_{i,V}),$$
where $y_{i,V}$ is the amount offered, $c$ is a level in the response variable, $F$ is the cumulative distribution function of the logistic density, $\tau_c$ is a cut point in the logistic and ($\tau_1 < \tau_2 < \ldots < \tau_{10}$). Only two respondents selected more than 500 GYD (one selected 600 and one 1000), so those responses have been collapsed into the 500 GYD category.

The number of times per year someone participated in *mayu* is a count variable with unequal mean and variance, necessitating the use of a negative binomial distribution (Cameron and Trivedi, 1998). This model uses a log link function to connect a linear predictor to the estimate for each individual and constrains the expected values to be above 0, as they must be if the outcome is a natural number. It estimates parameters for the following
$$Pr(Y_i = y_i) = \frac{\mu_i^{y_i} e^{-\mu_i}}{y_i!}$$
(the Poisson distribution) where
$$\mu_i = exp(\pi_{i,V} + \varepsilon_i)$$

And $exp(\varepsilon_i)$ is a gamma distributed variable with a mean of one and an independently estimated variance (Poch and Mannering, 1996). This allows us to model the mean of a Poisson distribution and the scale (or variation) of a Gamma distribution.

For each prediction, we examine two models: the model with random intercepts for each village but no fixed effects and the model with random intercepts for each village and fixed effects. We are interested in looking at if the variance structured at the village level captured in the village-intercept model can be explained by inclusion of the fixed effects. If the variance structured at the village level decreases from the inclusion of the fixed effects, we interpret that as showing how some of that village-level variance is structured. We also interpret the effect estimates in the full model.

We calculate the proportion of variance between communities in a null model and the proportion of the variance between communities in a model including the variables of interest using the intra-class correlation (ICC) for each of our models of interest. In general, the estimation of ICCs for generalized linear models is much less certain than for linear models. We should be wary of using ICCs produced using these methods to decide between models. Instead, we should limit our interpretation to the relative magnitude of the variation organized at different levels in different models.

To estimate the ICC for a multi-level ordered logistic regression, we follow the guidance of Snijders and Bosker (2011, section 17.4) and assume that the decisions represent a discrete realization of a continuous scale (which is a rather good assumption for our data), leading to a variance at the first level of 3.29. Calculation of the ICC is then $var_v/(var_v + 3.29)$, where $var_v$ is the variance at the village level. Importantly, the ICC estimated using this method can be much lower than the true ICC and give the impression of negligible within-cluster correlation (Grilli and Rampichini, 2012).

We implement one of the variations of equation 3.2 from Nakagawa and colleagues (2017) for the calculation of ICC in over-dispersed models such as the negative binomial. The ICC for this model is calculated as

$$\frac{var_V}{var_V + var_\varepsilon}$$

Where $var_v$ is the village-level variance and $var_\varepsilon$ is the observation-level variance. The observation-level variance is not to be confused with the variance associated with the observation-level random effect in an observation-level random effects model (indeed, this variance would only be a component of the observation-level variance). Nakagawa and colleagues (2017) show three methods for computing the observation-level variance, of which the trigamma function is the most accurate estimator of the observation-level variance for models with log link functions. As the negative binomial has a log link function, the observation-level variance will be calculated as

$$var_\varepsilon = \psi_1 \left[\frac{1}{\lambda} + \frac{1}{\theta}\right]^{-1}.$$

Where $\psi_1$ is the trigamma function, $\lambda$ is the mean of the outcome, and $\theta$ is the dispersion (sometimes called shape or size) parameter. While there are some complications with calculating $\lambda$, the mean of the response variable is a good approximation when sampling is balanced (as it mostly is in this case). In sum, ICC for these models is calculated as

$$\frac{var_V}{var_V + \psi_1 \left[\frac{1}{E[y]} + \frac{1}{\theta}\right]^{-1}}$$

## S7: Possible match of previously reported inconsistent effects on cooperation to predictions from multiplex overlap

Given the inconsistencies previously reported in the literature in predicting gameplay within communities, we reviewed the ethnographic materials to see if it were possible that variables in different communities would result in different levels of overlap. If, according to the ethnography, a variable seems to lead to higher (or lower) overlap and the data show lower (or higher) contributions in anonymous economic games, we classify that as supporting the multiplex overlap idea. If that relationship does not hold, we classify it as not supporting the multiplex overlap idea. As network overlap specifically was not collected, these associations are speculative, though ethnographically grounded.

*Support for effect of multiplex overlap on contributions in anonymous economic games*

- Hadza living in larger camps and with fewer siblings give more (Marlowe, 2004): This is predicated on the assumption that larger camps allow for less overlap, and that individuals with fewer coresident siblings in camp lack a core cooperative network of close kin. Both conditions would lead to individuals having lower network overlap.
- People in communities in Zimbabwe that had been forcibly resettled give more (Barr, 2004): This resettlement broke up existing networks (including kin groups), requiring people to form new connections in their new community, allowing for lower overlap.
- Au with more cocoa fields and increased wage labor give more (Tracer, 2004): This is predicated on the assumption that having more cocoa fields as well as engaging more often in wage labor leads one to have more interactions in "cocoa selling" and "wage labor" networks, reducing their importance in other networks.
- Orma with more income give more (Ensminger, 2004): income comes from engaging in a diversified set of strategies including wage labor, rock quarrying, and animal sales. This results in Orma with high income living in commercial area while Orma with low income living "at remote distances from towns and trade of any sort" (Ensminger, 2004, p. 361).
- Married Tsimane give less (Gurven et al., 2008): being married adds a core cooperative partner, not to mention possible core partners in close affinal kin, suggesting that these individuals have higher network overlap.
- Pahari Korwa with more connections outside the community give more while those with more sisters in their community give less (Lamba and Mace, 2011): connections outside the community increase the scope for decreased overlap while adult siblings are typically core cooperative partners around the world.

*Possible support for effect of multiplex overlap on contributions in anonymous economic games*

- Machiguenga who engage in large-scale cash cropping give more in anonymous settings than those who have ever engaged in wage labor as ethnographers report that large-scale cash cropping removes people from traditional networks more than having ever engaged in labor does (Henrich and Smith, 2004, p. 141)

- Selling more at market does not result in increased offers (Gurven et al., 2008) as it may be that Tsimane who are producing more crop for sale are still doing so using traditional networks.
- Tsimane who speak better Spanish giving less (Gurven et al., 2008) as people who speak better Spanish may be using that to consolidate power within their community.

*Contrary to predictions for effect of multiplex overlap on contributions in anonymous economic games*

- Pahari Korwa in larger communities took more from an anonymous public good (Lamba and Mace, 2011). This may be a result of the framing of the game, but further ethnographic information and research would be needed to explore why this is the case.

## References for SI